\documentclass[cits]{PoS}
\usepackage{epstopdf}

\title{Review of $\alpha_s$ determinations}

\ShortTitle{Review of $\alpha_s$ determinations}

\author{\speaker{Antonio Pich}          
        \\
        Departament de F\'{\i}sica Te\`orica, IFIC, Universitat de Val\`encia -- CSIC, Edifici d'Instituts de Paterna, Apt. Correus 22085, E-46071, Val\`encia, Spain\\
        E-mail: \email{antonio.pich@ific.uv.es}}

\abstract{The present knowledge on the strong coupling is briefly summarized. The most precise determinations of $\alpha_s$, at different energies, are reviewed and compared at the Z mass scale, using the predicted QCD running. The impressive agreement achieved between experimental measurements and theoretical predictions constitutes a beautiful and very significant test of Asymptotic Freedom, establishing QCD as the fundamental theory of the strong interaction. The world average value of the strong coupling is found to be
$\alpha_s(M_Z^2)= 0.1186 \pm 0.0007$.}

\FullConference{Xth Quark Confinement and the Hadron Spectrum\\
		 8--12 October 2012\\ TUM Campus Garching, Munich, Germany}

\begin{document}

\section{Introduction}

In the massless quark limit, Quantum Chromodynamics (QCD) has only one free parameter: the strong coupling $\alpha_s$. All strong interaction phenomena should be described in terms of this single input. The measurements of $\alpha_s$ at different processes and at different mass scales provide then a crucial test of QCD. Obviously, the test should be restricted to those processes where perturbative techniques are reliable. Moreover, the same definition of $\alpha_s$ should be taken everywhere; the $\overline{\mathrm{MS}}$ scheme is usually adopted as the standard convention. Since the running coupling is a function of energy, one can either compare the different determinations at the different scales where they are measured, checking in this way the predicted momentum dependence of the coupling, or use this prediction to bring all measurements to a common reference scale where they are compared. Nowadays, the Z mass scale is conventionally chosen for such a comparison. In order to assess the significance of the test, it is very important to have a good understanding of the uncertainties associated with the different measurements. This is not easy because small non-perturbative effects can be present in many observables. In addition, some quantities have been computed to a very good perturbative accuracy,  next-to-next-to-leading order (NNLO) or even next-to-next-to-next-to-leading order (N${}^3$LO), while others are only known at the leading (LO) or next-to-leading order (NLO); the resulting values of $\alpha_s$ refer then to different perturbative approximations. The theoretical predictions are also affected by the expected asymptotic ({\it i.e.}, non-convergent) behaviour of the perturbative series in powers of $\alpha_s$. Although this is a common problem of Quantum Field Theories, it is probably more severe in QCD because the coupling is rather large (at usual energies).

The most precise determinations of the strong coupling were reviewed in detail in Refs.~\cite{Bethke:2009jm,Bethke:2011tr} and have been recently updated \cite{BDS:12} in the 2012 review of particle physics \cite{Beringer:1900zz}. I will heavily use the information provided in these references, adding my personal biases and updating the summary with the most recent developments.

\section{QCD running and threshold matching}

The scale dependence of the QCD coupling is governed by the
renormalization group equation
\begin{equation}
\mu\,\frac{d\alpha_s(\mu^2)}{d\mu}\; =\; \alpha_s(\mu^2)\,\beta(\alpha_s)\, ,
\qquad\qquad\qquad
\beta(\alpha_s)\; =\;\sum_{n=1}\, \beta_n\,\left(\frac{\alpha_s}{\pi}\right)^n\, ,
\end{equation}
where the $\overline{\mathrm{MS}}$ $\beta$ function is already known to the fourth order \cite{vanRitbergen:1997va,Czakon:2004bu} ($\zeta_3 = 1.202056903 \ldots$):
\begin{eqnarray}
\beta_1 & =& \frac{1}{3}\, N_f - \frac{11}{2}\, ,
\qquad\quad
\beta_2\, =\, -{51\over 4} + {19\over 12} \, N_f \, ,
\qquad\quad
\beta_3 \, = \, {1\over 64}\left[ -2857 + {5033\over 9} \, N_f
- {325\over 27} \, N_f^2 \right] \, ,
\nonumber\\
\beta_4 &=&  \frac{-1}{128}\, \left[
\left( \frac{149753}{6}+3564 \,\zeta_3 \right)
-\left( \frac{1078361}{162}+\frac{6508}{27}\,\zeta_3 \right) N_f
 \right.
\nonumber \\ & & \quad\;\;\:\left.\mbox{}
+ \left( \frac{50065}{162}+\frac{6472}{81}\,\zeta_3 \right) N_f^2
+\frac{1093}{729}\, N_f^3 \right]~.
\end{eqnarray}
The explicit dependence with the number of quark flavours $N_f$ shows the need to properly match the different QCD${}_{N_f}$ effective theories when crossing the corresponding quark thresholds; {\it i.e.}, the value of $\alpha_s$ depends on the number of `active' quarks considered. If one quark flavour is removed from the Lagrangian (its mass taken to be heavy enough to decouple), the couplings of the original theory with $N_f+1$ flavours and the resulting effective one with $N_f$ flavours are related through the matching condition
\begin{equation}
\alpha_s^{(N_f)}(\mu^2)\; =\; \alpha_s^{(N_f+1)}(\mu^2)\;\left\{ 1 +
\sum_{k=1}^\infty\, \sum_{n=0}^k\; c_{kn}\; \left[\frac{\alpha_s^{(N_f+1)}(\mu^2)}{\pi}\right]^k\,\log^n{\left[
\frac{\mu^2}{M^2(\mu^2)}\right]}\right\}\, ,
\end{equation}
where $M(\mu^2)$ in the running mass of the heavy quark.
The coefficients $c_{kn}$ are known up to four loops ($k\le 4$) \cite{Schroder:2005hy,Chetyrkin:2005ia}:
\begin{eqnarray}
c_{10} & = & 0\, ,\qquad\quad c_{11}\; =\; -\frac{1}{6}\, , \qquad\quad
c_{20}\; =\; \frac{11}{72}\, , \qquad\quad c_{21}\; =\; -\frac{11}{24}\, , \qquad\quad
c_{22}\; =\; \frac{1}{36}\, ,
\nonumber\\
c_{30}& =& \frac{564731}{124416}-\frac{82043}{27648}\;\zeta_3
-\frac{2633}{31104}\; N_f\, , \qquad
c_{31}\; =\; -\frac{955}{576}+ \frac{67}{576}\; N_f\, , \qquad
c_{32}\; =\; \frac{53}{576}-\frac{1}{36}\; N_f\, ,
\nonumber\\
c_{33}& =& -\frac{1}{216}\, ,\qquad\quad
c_{40}\; =\; 5.17035 - 1.00993\; N_f - 0.0219784\; N_f^2\, ,\qquad\quad
c_{44}\; =\;\frac{1}{1296}\, ,
\nonumber\\
c_{41}& =& \frac{7391699}{746496}-\frac{2529743}{165888}\;\zeta_3
-\left(\frac{110341}{373248}-\frac{110779}{82944}\;\zeta_3\right)\, N_f +\frac{6865}{186624}\; N_f^2\, ,
\nonumber\\
c_{42}& =& \frac{2177}{3456}-\frac{1483}{10368}\; N_f -\frac{77}{20736}\; N_f^2\, , \qquad\quad\quad
c_{43}\; =\; -\frac{1883}{10368}-\frac{127}{5184}\; N_f +\frac{1}{324}\; N_f^2\, .
\end{eqnarray}
Since $c_{10}=0$, the discontinuity in $\alpha_s$ is small for $\mu=M$, but a larger effect is found with other choices
of the matching point. Physics should of course not depend on this choice and one should carefully evaluate the theoretical uncertainties, allowing $\mu$ to vary within a reasonable range.

\begin{figure}
\includegraphics[width=.48\textwidth]{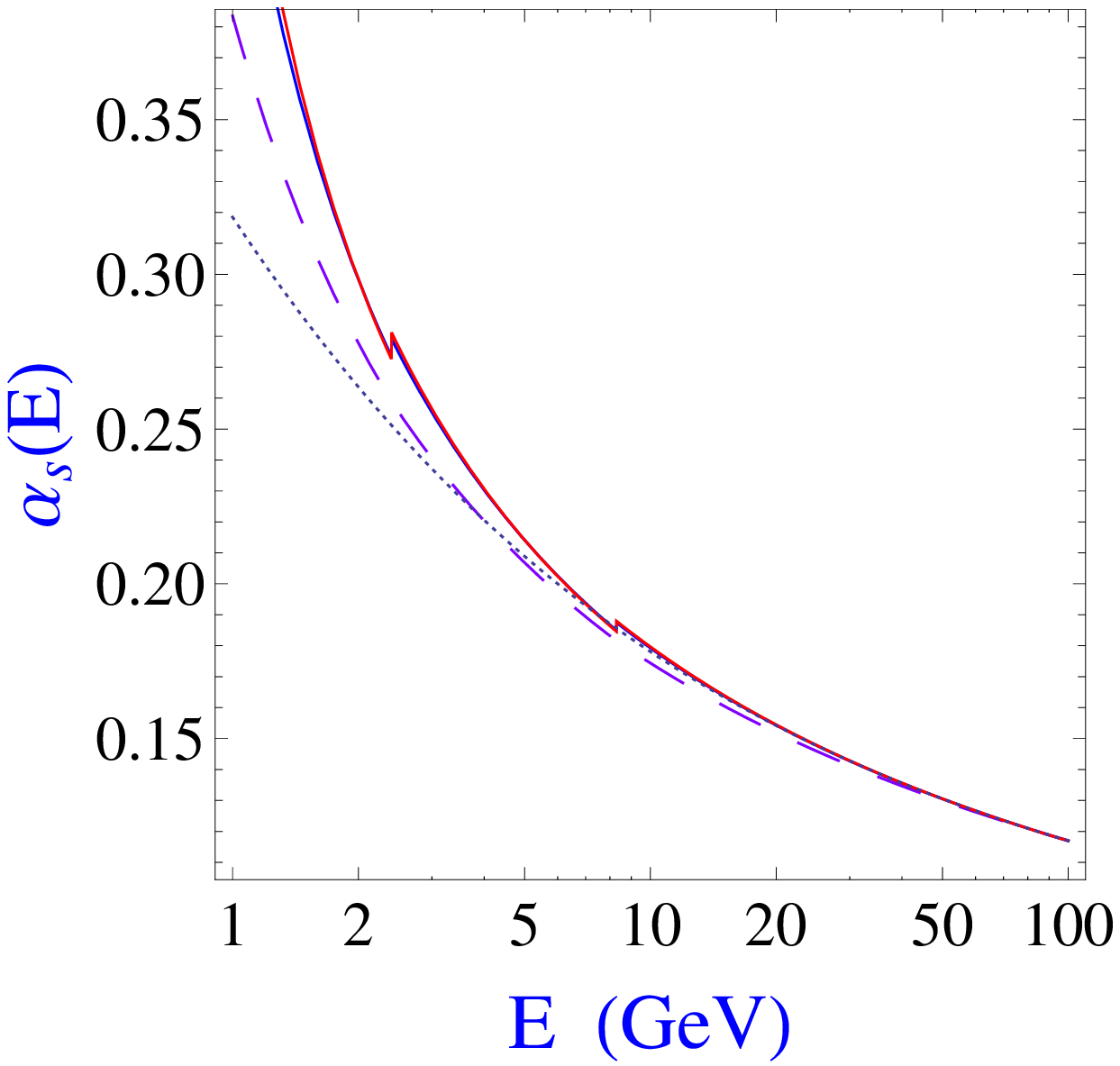}
\hskip .5cm
\includegraphics[width=.48\textwidth]{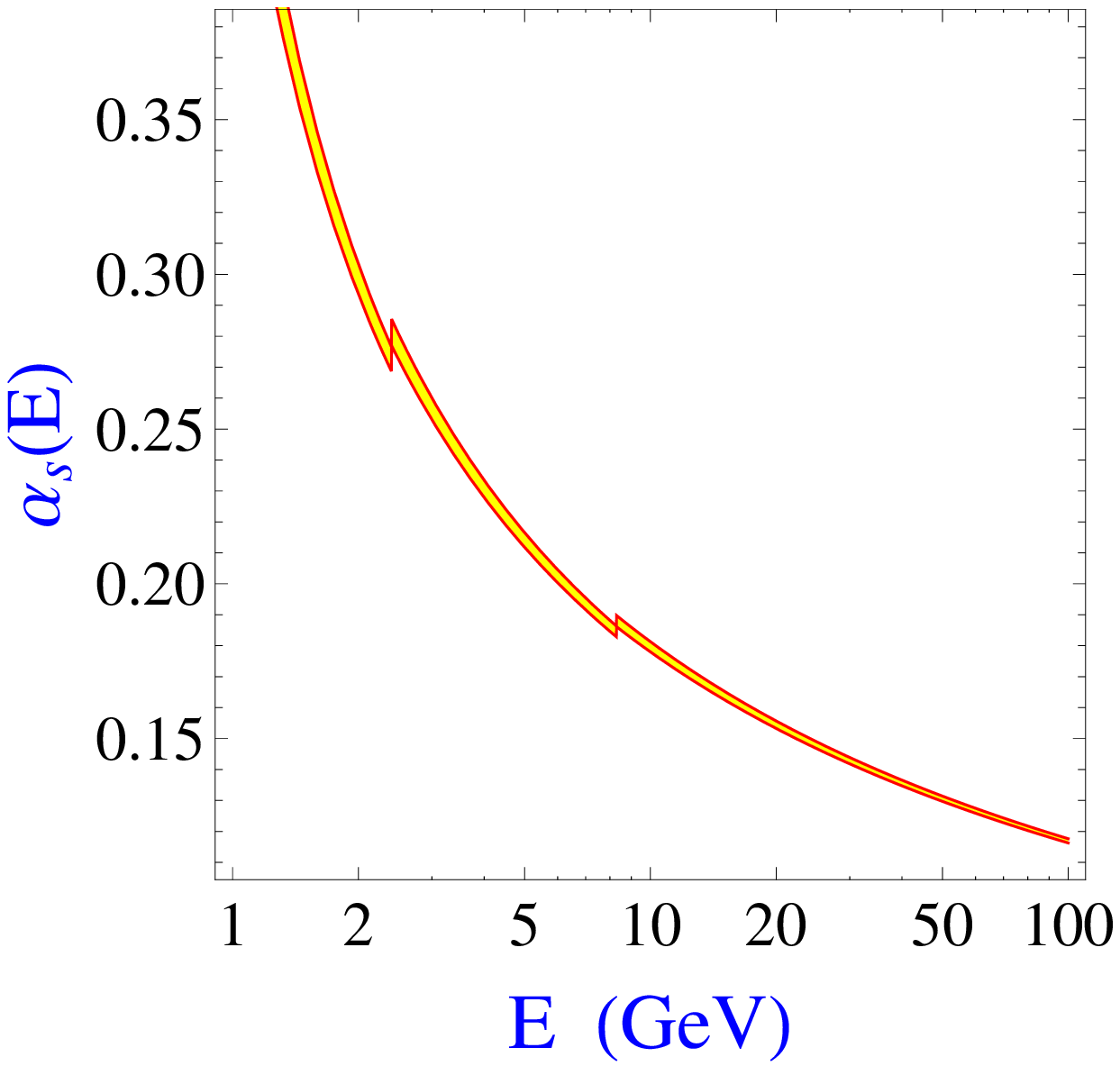}
\caption{Scale dependence of $\alpha_s$ at one (dashed), two (continuous, blue) and four (continuous, red) loops (left). The dotted line is the 4-loop FOPT approximation
in powers of $\alpha_s(M_Z^2) = 0.1186$.
The right plot shows the 4-loop error evolution taking $\alpha_s(M_Z^2) = 0.1186\pm 0.0007$; the uncertainty increases at lower energies.}
\label{fig:running}
\end{figure}

The 4-loop running coupling allows us to perform a resummation of N${}^3$LO logarithms, {\it i.e.}, corrections of the form
$\alpha_s(\mu^2)^{n+4} \log^n{(Q^2/\mu^2)}$, which are very sizeable when the difference of scales is large. This is illustrated in Fig.~\ref{fig:running}, which
shows the scale dependence of $\alpha_s$ over a wide range of energies, at different levels of approximation. The one-loop $\beta$-function resummation does already a very good job, and the accuracy rapidly improves at higher orders. However, a naive fixed-order perturbative expansion (FOPT) in powers of $\alpha_s(M_Z^2)$ does not approximate well the result at low energies, even at four loops, because $\log{(E/M_Z)}$ is large.

\section{Inclusive observables:\ $\mathbf{\sigma(e^+e^-\to\mathrm{\mathbf{hadrons}})}$ \ and \ $\boldmath \Gamma(Z\to\mathrm{\bf hadrons})$}

The inclusive production of hadrons mediated by $\gamma$,
$Z$ and $W^\pm$ interactions proceeds through the vector
$\, V^{\mu}_{ij} = \bar{\psi}_j \gamma^{\mu} \psi_i \, $
and axial-vector
$\, A^{\mu}_{ij} = \bar{\psi}_j \gamma^{\mu} \gamma_5 \psi_i \,$
colour-singlet quark currents ($i,j=u,d,s\ldots$).
The theoretical analysis involves the two-point correlation functions
\begin{equation}\label{eq:pi_v2}
\Pi^{\mu \nu}_{ij,J}(q)\; \equiv\;
 i \int d^4x \;\, \mathrm{e}^{iqx}\,
\langle 0|T(J^{\mu}_{ij}(x)\, J^{\nu}_{ij}(0)^\dagger)|0\rangle
\; =\;
\left( -g^{\mu\nu} q^2 + q^{\mu} q^{\nu}\right) \, \Pi_{ij,J}^{(1)}(q^2)
 +   q^{\mu} q^{\nu} \, \Pi_{ij,J}^{(0)}(q^2) \, ,
\end{equation}
where the superscript $(0,1)$ denotes the angular momentum in the hadronic rest frame.
For massless quarks, $s\,  \Pi_{ij,J}^{(0)}(s)= \mathrm{constant}$ (there is a non-perturbative Goldstone-pole contribution to $\Pi_{ij,A}^{(0)}$ at $s=0$, which cancels in $\Pi_{ij,A}^{(0+1)}$). These correlators are already known to $\mathcal{O}(\alpha_s^4)$ \cite{Baikov:2008jh,Baikov:2012er}. In the so-called non-singlet contributions ($i\not= j$), one quark loop connects the two external currents. If quark masses are neglected, these contributions are identical for the vector and axial-vector correlators:\
$\Pi(s)\equiv \Pi_{i\not=j,V}^{(0+1)}(s)=\Pi_{i\not=j,A}^{(0+1)}(s)$.
They are conveniently parametrized through the Euclidean Adler function ($Q^2=-q^2$)
\begin{equation}\label{eq:Adler}
D(Q^2)\; \equiv\; -Q^2\frac{d}{dQ^2}\Pi(Q^2)\; =\; \frac{N_C}{12\pi^2}\;\left\{
1 + \sum_{n=1}\; K_n
\left( {\alpha_s(Q^2)\over \pi}\right)^n\right\}
\, ,
\end{equation}
with $N_C=3$ the number of quark colours.
The known ($n\le 4)$ coefficients take the values \cite{Baikov:2008jh,Gorishnii:1990vf,Surguladze:1990tg}:
\begin{eqnarray}
K_1 &=& 1\, ,
\qquad
K_2\; =\; 1.98571 -0.115295\; N_f\, ,
\qquad
K_3\; =\; 18.2427 - 4.21585\; N_f + 0.0862069\; N_f^2\, ,
\nonumber\\
K_4 &=& 135.792 - 34.4402\; N_f + 1.87525\; N_f^2 - 0.0100928\; N_f^3\, .
\end{eqnarray}
For neutral currents ($i=j$) there are additional singlet contributions where
each current couples to a different quark loop. Owing to the gluon quantum numbers
(colour octet and $J^{PC}=1^{--}$), these contributions start at $\mathcal{O}(\alpha_s^3)$ and $\mathcal{O}(\alpha_s^2)$, respectively, for the vector and axial-vector currents:
\begin{eqnarray}
\delta^\mathrm{S} D_V(Q^2)\; =\; \frac{N_C}{12\pi^2}\;\sum_{n=3}\; d^{V}_n
\left( {\alpha_s(Q^2)\over \pi}\right)^n\, ,
\qquad\quad
\delta^\mathrm{S} D_A(Q^2)\; =\; \frac{N_C}{12\pi^2}\;\sum_{n=2}\; d^{A}_n
\left( {\alpha_s(Q^2)\over \pi}\right)^n\, .\quad
\end{eqnarray}
The vector-current coefficients are
$d_3^V = -0.41318$ and $d_4^V = -5.94225 + 0.191628\; N_f$ \cite{Baikov:2012er}.

The electromagnetic hadronic production in $e^+e^-$ annihilation is given by
\begin{eqnarray}\label{eq:R_ee}
R_{e^+e^-}(s)& \equiv&
\frac{\sigma(e^+e^-\to \mathrm{hadrons})}{\sigma(e^+e^-\to\mu^+\mu^-)}\;
=\; 12 \pi \;\left\{ \sum_f Q_f^2\;\, \mathrm{Im} \Pi(s)
+ \left(\sum_f Q_f\right)^{\! 2}\; \mathrm{Im}\, \delta^\mathrm{S}\Pi_V(s)\right\}
\nonumber\\
& =& \sum_{f} Q_f^2\; N_C \; \left\{ 1 +
\sum_{n\geq 1} F_n \left({\alpha_s(s)\over\pi}\right)^{\! n} \right\}
\; + \; \mathcal{O}\left(\frac{m_q^2}{s},\frac{\Lambda^4}{s^2}\right)\, .
\end{eqnarray}
The sum over quark electric charges of different signs strongly suppresses the singlet contribution, which has been included as a small correction to the coefficients $F_{n\ge 3}$. For $N_f=5$ flavours, one gets\ $F_1 = 1$, $F_2 = 1.4092$,
$F_3 = -12.805$ and $F_4 = -80.434$ \cite{Baikov:2012er}.

The calculated perturbative series is actually an expansion in powers of $\alpha_s(\mu^2)$ with coefficients containing a polynomial dependence on $\log{(s/\mu^2)}$. In Eq.~(\ref{eq:R_ee}), these logarithms have been resummed into the running coupling by taking $\mu^2=s$. The physical ratio $R_{e^+e^-}(s)$ is of course independent of the renormalization scale $\mu$, but the truncated series contains a residual $\mu$-dependence of $\mathcal{O}(\alpha_s^{N+1})$, where $N=4$ is the last included term, which must be taken into account in the theoretical uncertainty. Since non-perturbative corrections are suppressed by $\Lambda^4/s^2$ (the gauge-invariant operators contributing to the current correlators have dimensions $D\ge 4$), at high energies one can perform a N${}^3$LO determination of $\alpha_s(s)$. Unfortunately, the experimental uncertainties are large.

A much more accurate determination can be obtained from the precise experimental measurement of the hadronic $Z$ decay rate. The electroweak neutral current
$J^\mu_Z = \sum_f (v_f V_{ff}^\mu + a_f A_{ff}^\mu)$ involves the vector and axial-vector currents, weighted with the corresponding $Z$ couplings. Since $a_f = 2 I_f$, the singlet axial contributions of the two members of a weak isospin doublet cancel each other for equal quark masses; however, the large top mass generates very important
singlet axial corrections which start at $\mathcal{O}(\alpha_s^2)$. The resulting QCD series takes the form
\begin{equation}\label{eq:RZ}
R_Z\; \equiv\; \frac{\Gamma(Z\to\mathrm{hadrons})}{\Gamma(Z\to e^+e^-)}\; =\; R_Z^{\mathrm{EW}} \; N_C\;\left\{
1 + \sum_{n=1}\; \tilde F_n
\left( {\alpha_s(M_Z^2)\over \pi}\right)^n\right\}
\, ,
\end{equation}
with $\tilde F_1 = 1$, $\tilde F_2 = 0.76264$,
$\tilde F_3 = -15.490$ and $\tilde F_4 = -68.241$ \cite{Baikov:2012er}.
Taking properly into account the electroweak corrections, the ratio $R_Z$ is included in the global fit to electroweak precision data, which results in a quite accurate determination of $\alpha_s(M_Z^2)$. Note that this assumes the validity of the electroweak Standard Model with the minimal Higgs mechanism. The PDG quotes the result~\cite{Beringer:1900zz}
\begin{equation}\label{eq:alpha_Z}
\alpha_s^{(N_f=5)}(M_Z^2)\; \equiv\; \alpha_s(M_Z^2)\; =\; 0.1197\pm 0.0028\, .
\end{equation}

\section{The hadronic $\tau$ decay width}

The inclusive hadronic decay of the $\tau$ lepton provides a very clean way to determine $\alpha_s$ at low energies \cite{Narison:1988ni}. The QCD correlation function of two left-handed $W$ currents receives only non-singlet contributions. Restricting the analysis to the dominant Cabibbo-allowed decay width,
\begin{eqnarray}\label{eq:R_tau}
R_{\tau,V+A} &\equiv &\frac{ \Gamma [\tau^- \to \nu_\tau +\mathrm{hadrons}\, (S=0)]}{ \Gamma [\tau^- \to \nu_\tau e^- {\bar \nu}_e]}
\nonumber\\
& = & 12 \pi\; |V_{ud}|^2 \int^{m_\tau^2}_0 {ds \over m_\tau^2 } \,
 \left(1-{s \over m_\tau^2}\right)^2
\biggl[ \left(1 + 2 {s \over m_\tau^2}\right)
 \mbox{\rm Im} \Pi^{(1)}_{ud,V+A}(s)
 + \mbox{\rm Im} \Pi^{(0)}_{ud,V+A}(s) \biggr]  .\qquad
\end{eqnarray}
Although at low $s$ the integrand cannot be predicted from first principles, the integral itself can be calculated systematically by exploiting
the analytic properties of the current correlators. They are analytic
functions of $s$ except along the positive real $s$-axis, where their
imaginary parts have discontinuities.
$R_{\tau,V+A}$ can then be written as a contour integral
in the complex $s$-plane running
counter-clockwise around the circle $|s|=m_\tau^2$ \cite{Narison:1988ni,Braaten:1988hc,Braaten:1991qm}:
\begin{equation}\label{eq:circle}
 R_{\tau,V+A} \; =\;
6 \pi i\; |V_{ud}|^2\; \oint_{|s|=m_\tau^2} {ds \over m_\tau^2} \,
 \left(1 - {s \over m_\tau^2}\right)^2
\left[ \left(1 + 2 {s \over m_\tau^2}\right) \Pi^{(0+1)}_{ud,V+A}(s)
         - 2 {s \over m_\tau^2} \Pi^{(0)}_{ud,V+A}(s) \right]  .
\end{equation}
In the chiral limit ($m_{u,d} = 0$), only the correlator
$\Pi^{(0+1)}_{ud,V+A}(s)$ contributes.
Using the Operator Product Expansion (OPE),
$\Pi^{(0+1)}(s) = \sum_{D} C_D/ (-s)^{D/2}$,
to evaluate the contour integral, $R_{\tau,V+A}$
can be expressed as an expansion in powers of $1/m_\tau^2$
\cite{Braaten:1991qm}.
The theoretical prediction can be written as
\begin{equation}\label{eq:Rv+a}
 R_{\tau,V+A} \; =\; N_C\; |V_{ud}|^2\; S_{\mathrm{EW}}\; \left\{ 1 +
 \delta_{\mathrm{P}} + \delta_{\mathrm{NP}} \right\} \, ,
\end{equation}
where $S_{\mathrm{EW}}=1.0201\pm 0.0003$ contains the
electroweak radiative corrections \cite{Marciano:1988vm,Braaten:1990ef,Erler:2002mv}.
The non-perturbative contributions are suppressed
by six powers of the $\tau$ mass \cite{Braaten:1991qm}\footnote{
Since $(1 - x)^2 (1 + 2 x) = 1 - 3 x^2 + 2x^3$, 
the only non-perturbative contributions to the circle integration originate from operators with dimensions $D=6$ and 8 (up to tiny logarithmic running corrections).}
and can be extracted from the invariant-mass distribution of the final
hadrons \cite{Le Diberder:1992fr,Schael:2005am,Ackerstaff:1998yj,Coan:1995nk}.
The presently most complete and precise experimental analysis, performed with
the ALEPH data, obtains $\delta_{\mathrm{NP}} = -0.0059\pm 0.0014$ \cite{Davier:2008sk}.
Quark mass effects  \cite{Braaten:1991qm,Pich:1998yn,Baikov:2004tk}
amount to a negligible correction smaller than $10^{-4}$.

The dominant correction ($\sim 20\%$) is the perturbative QCD
contribution ($N_f=3$)
%
\begin{equation}\label{eq:r_k_exp}
\delta_{\mathrm{P}} \; =\;
\sum_{n=1}  K_n \; A^{(n)}(\alpha_s)
\; =\; a_\tau + 5.20\; a_\tau^2 + 26\; a_\tau^3 + 127\; a_\tau^4
 + \cdots\, ,
\end{equation}
where the contour integrals \cite{Le Diberder:1992te}
\begin{equation}\label{eq:a_xi}
A^{(n)}(\alpha_s) \, = \,
 {1\over 2 \pi i}\;\oint_{|s| = m_\tau^2} {ds \over s} \;
  \left({\alpha_s(-s)\over\pi}\right)^n
 \left( 1 - 2 {s \over m_\tau^2} + 2 {s^3 \over m_\tau^6}
         - {s^4 \over  m_\tau^8} \right)
\end{equation}
only depend on
$a_\tau\equiv\alpha_s(m_\tau^2)/\pi$.
The main uncertainty originates in the treatment of higher-order corrections
\cite{Baikov:2008jh,Davier:2008sk,Pich:2011bb,Beneke:2008ad,Caprini:2011ya,Abbas:2012fi,Abbas:2012py,Groote:2012jq,Maltman:2008nf,Menke:2009vg,Narison:2009vy,Cvetic:2010ut}.
The FOPT expansion in powers of $a_\tau$, shown in the r.h.s of Eq.~(\ref{eq:r_k_exp}), has
coefficients much larger than $K_n$
because the long running of $\alpha_s$ along the circle generates large logarithms, $\log{(-s/m_\tau^2)}= i\phi$ ($\phi\in [-\pi,\pi]$), giving rise to a non-convergent series. These large corrections can be resummed to all orders, using the renormalization-group $\beta$-function equation to compute exactly
(up to unknown $\beta_{n>4}$ contributions) the functions $A^{(n)}(\alpha_s)$. One obtains then a well-behaved expansion, known as contour-improved perturbation theory (CIPT) \cite{Le Diberder:1992te,Pivovarov:1991rh}, which results in smaller values for $\delta_{\mathrm{P}}$ than the FOPT approach.
However, assuming that the known $n\le 4$ terms of the Adler series are already governed by low-lying infrared renormalons ({\it i.e.}, already sensitive to the asymptotic series regime), it has been argued that CIPT could miss cancelations induced by the renormalonic behaviour \cite{Beneke:2008ad}; in that case, FOPT could approach faster the Borel summation of the full renormalon series. Making optimal conformal mappings in the Borel plane and properly implementing the CIPT procedure within the Borel transform, one also obtains numerical results close to the FOPT value \cite{Caprini:2011ya,Abbas:2012fi}.

The present experimental value $R_{\tau,V+A} = 3.4671\pm 0.0084$ \cite{HFAG} implies $\delta_P = 0.1995\pm 0.0033$. Using CIPT one gets $\alpha_s(m_\tau^2) = 0.339\pm 0.013$, while FOPT would give
$\alpha_s(m_\tau^2) = 0.318\pm 0.014$  \cite{Pich:2011bb}. Combining the two results, but keeping conservatively the smallest error, we get
\begin{equation}\label{eq:alpha-result}
\alpha_s^{(N_f=3)}(m_\tau^2)\; =\; 0.329 \pm 0.013\, ,
\qquad\qquad\quad
\alpha_s(M_Z^2)\;  =\;  0.1198\pm 0.0015\, .
\end{equation}
Although $\alpha_s^{(N_f=3)}(m_\tau^2)$ is significantly larger ($16\,\sigma$) than the result obtained
from the $Z$ hadronic width in Eq.~(\ref{eq:alpha_Z}), after evolution up to the scale $M_Z$ there is an
excellent agreement between the two measurements. These two determinations
provide a beautiful test of the predicted QCD running;
{\it i.e.}, a very significant experimental verification of {\it asymptotic freedom}.

At the presently achieved accuracy, a better experimental assessment of $\delta_{\mathrm{NP}}$ would be welcome,
which requires a more precise determination of the inclusive hadronic distribution. This would also allow for an investigation of uncertainties related with the use of the OPE near the time-like axis
(duality violations) \cite{Davier:2008sk,DescotesGenon:2010cr,Cata:2008ru}; in $R_\tau$ they are heavily suppressed by the presence in (\ref{eq:circle}) of a double zero at $s=m_\tau^2$ \cite{Braaten:1991qm}, but they could be more relevant for other moments of the hadronic invariant-mass distribution. A recent fit to rescaled OPAL data, with moments chosen to maximize duality violations, finds
$\delta_{\mathrm{NP}} = -0.003\pm 0.012$ and $\alpha_s(m_\tau^2) = 0.333\pm 0.018$ (CIPT+FOPT)
\cite{Boito:2012cr}, in agreement but less precise than the result obtained from the ALEPH
invariant-mass distribution,
$\delta_{\mathrm{NP}} = -0.0059\pm 0.0014$ and $\alpha_s(m_\tau^2) = 0.344\pm 0.009$ (CIPT) \cite{Davier:2008sk}.


\section{Lattice determination}

Lattice simulations determine the strong coupling by measuring various short-distance quantities (non-perturbatively),
through a numerical evaluation of the QCD functional integral,
and comparing the results with the corresponding perturbative expansions in powers of $\alpha_s$,
using lattice QCD perturbation theory which includes lattice-spacing artifacts.
Modern simulations are done with $2+1$ flavours of sea quarks (one tuned to the strange quark
and the other two taken with masses as small as possible for up and down) and have a NNLO perturbative accuracy.
At least one dimensionful physical quantity is needed to convert from lattice units to GeV, {\it i.e.}, to fix the scale at which $\alpha_s$ is measured.

The HPQCD collaboration has extracted the coupling, following two different approaches with different systematics.
They use staggered fermions and the lattice spacing is determined from a wide variety of physical quantities.
From 22 different simulations of small Wilson loops they obtain the value $\alpha_s(M_Z^2) = 0.1184\pm 0.0006$ \cite{McNeile:2010ji}.
In the second approach, they measure four moments of the correlation function of two heavy-quark currents,
at 8 different values of the heavy-quark mass between $m_c$ and $m_b$ and 5 different lattice spacings, and obtain
$\alpha_s(M_Z^2) = 0.1183\pm 0.0007$~\cite{McNeile:2010ji}.
An independent perturbative analysis, using the results of a previous HPQCD-UKQCD simulation \cite{Mason:2005zx} already superseded by the new data,
finds a slightly larger value $\alpha_s(M_Z^2) = 0.1192\pm 0.0011$~\cite{Maltman:2008bx}.

The PACS-CS collaboration adopts the so-called Schr\"odinger functional scheme to carry out a non-perturbative running of the coupling, from the low energy region used to introduce the physical scale to high energies where the matching to the $\overline{\mathrm{MS}}$ scheme is performed.
The simulations are done with the Iwasaki gauge action and the non-perturbatively improved Wilson-fermion action with the clover term. They quote the result
$\alpha_s(M_Z^2) = 0.1205\pm 0.0008\pm 0.0005\,{}^{+\, 0.0000}_{-\, 0.0017}$ \cite{Aoki:2009tf}.
The Schr\"odinger functional scheme has been also used by the ALPHA collaboration to study the
running of $\alpha_s$ with $N_f=0,2$ dynamical flavours, but the results with $N_f=4$ are not yet complete~\cite{DellaMorte:2004bc}.

A numerical simulation of the Adler function, performed with dynamical overlap fermions by the JLQCD collaboration, finds
$\alpha_s(M_Z^2) = 0.1181\pm 0.0003\,{}^{+\, 0.0014}_{-\, 0.0012}$ \cite{Shintani:2010ph}. The uncertainty is dominated by the lattice spacing due to the relatively small and coarse lattice used.

More recently, the ETM collaboration has simulated the ghost-gluon vertex, including a dynamical charm quark ($N_f = 2+1+1$),
and has studied the running of $\alpha_s$ over a wide momentum window between 1.7 and 6.8 GeV.
Fitting the data with a MOM-like Taylor coupling plus a $1/p^6$ non-perturbative term, they obtain 
$\alpha_s^{(N_f=4)}(m_\tau^2) = 0.339\pm 0.013$, in good agreement with the $\tau$ determination in Eq.~(\ref{eq:alpha-result}). They finally quote
$\alpha_s(M_Z^2) = 0.1200\pm 0.0014$ \cite{Blossier:2012ef}.
An analogous MOM-like study of the triple-gluon vertex with domain-wall fermions and the Iwasaki gauge action has been
presented by RBC/UKQCD at the last Lattice conference \cite{RBC:12}, with the preliminary value
$\alpha_s(M_Z^2) = 0.1202\pm 0.0011\pm 0.0002\pm 0.0039\pm 0.0060$, where the last and dominant uncertainty accounts for
the 5\% estimated finite-volume error.

The energy between a static quark and a static antiquark separated by a distance $r$ has been also calculated in the lattice,
with $2+1$ flavours, combining a tree-level improved gauge action with a highly-improved staggered quark action \cite{Bazavov:2011nk}. The simulations have been done over a wide range of gauge couplings, corresponding to a lattice spacing
$1.909/r_0 \le a^{-1} \le 6.991/r_0$, with $r_0 = 0.468\pm 0.004$~fm.
The short-distance part of the static energy can be computed perturbatively and it is nowadays known at next-to-next-to-next-to-leading logarithmic (N${}^3$LL) accuracy, {\it i.e.}, including terms up to order $\alpha_s^{4+n}\log^n{\alpha_s}$ with $n\ge 0$ \cite{Brambilla:2010pp,Pineda:2011db,Smirnov:2009fh,Anzai:2009tm}
(the $\log{\alpha_s}$ terms are generated by virtual emissions of ultrasoft gluons).
Assuming that QCD perturbation theory (after canceling a renormalon contribution) is enough to describe the lattice data at distances $r<0.5\; r_0$, a fit to the lattice results leads to $r_0\;\Lambda_{\overline{\mathrm{MS}}} = 0.70\pm 0.07$, or equivalently,
$\alpha_s^{(N_f=3)}(1.5\:\mathrm{GeV})\; =\; 0.326 \pm 0.0019$
and
$\alpha_s(M_Z^2)\; =\; 0.1156 \, {}^{+\, 0.0021}_{-\, 0.0022}$ \cite{Bazavov:2012ka}.
Since the present lattice data are not precise enough to profit from the N${}^3$LL calculation, the determination has been obtained at three loop plus leading ultrasoft logarithmic resummation accuracy.

\begin{figure}[t]\centering
\includegraphics[width=.6\textwidth]{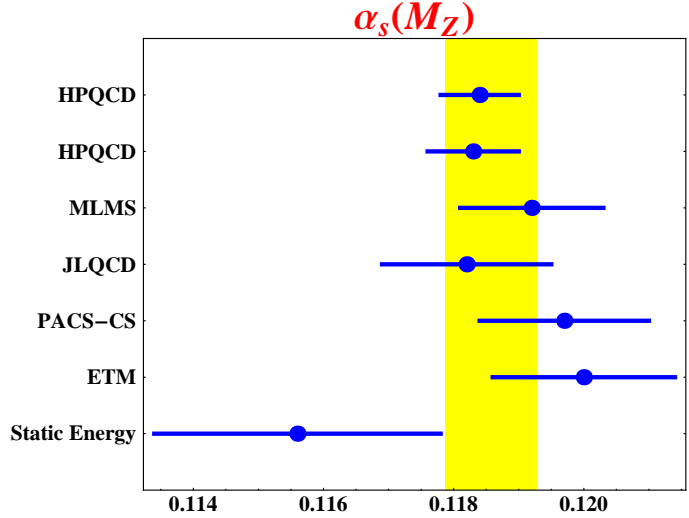}
\caption{Lattice determinations of $\alpha_s(M_Z^2)$. The yellow band shows the average in (\protect\ref{eq:lattice}).}
\label{fig:Lattice-pre-average}
\end{figure}

The results of Refs.~\cite{McNeile:2010ji,Aoki:2009tf,Shintani:2010ph,Blossier:2012ef,Bazavov:2012ka} are largely independent.
Following the prescription advocated in Ref.~\cite{BDS:12}, we take
a weighted average with a correlated error so that the overall
$\chi^2$ equals unity per degree of freedom. This  gives
\begin{equation}\label{eq:lattice}
\alpha_s(M_Z^2)\; =\; 0.1186 \pm 0.0007\, .
\end{equation}

\section{Quarkonium}

The bound states of a heavy quark and a heavy antiquark can be rigorously described with
non-relativistic QCD (NRQCD) techniques, through a combined expansion in powers of $\alpha_s$ and the heavy-quark velocity $v$. The strong coupling can then be extracted from the ratio \cite{Besson:2005jv,GarciaiTormo:2005ch}
\begin{equation}\label{eq:R_quarkonium}
R_\gamma\;\equiv\;\frac{\Gamma[\,\Upsilon(1S)\to\gamma +\mathrm{hadrons}]}{\Gamma[\,\Upsilon(1S)\to \mathrm{hadrons}]}
\; =\; 0.0245\pm 0.0001\pm 0.0013\, .
\end{equation}
Including the color-octet contributions, which are important in the upper end-point region of the photon spectrum, one gets a good description of the photon energy distribution \cite{GarciaiTormo:2005ch}. In order to extract $\alpha_s$, a sensible determination of the NRQCD hadronic matrix elements is needed. Using a combination of lattice and continuum inputs, the most recent analysis obtains \cite{Brambilla:2007cz}
\begin{equation}\label{eq:quarkonium}
\alpha_s(M_{\Upsilon(1S)}^2)\; =\; 0.184 \,{}^{+\, 0.015}_{-\, 0.014}\, ,
\qquad\qquad\qquad
\alpha_s(M_Z^2)\; =\; 0.119 \,{}^{+\, 0.006}_{-\, 0.005}\, .
\end{equation}
Since the accuracy of this determination is only of NLO in $\alpha_s(m_b^2)$ and $v^2$, it will not be taken into account in our final average. However, the result nicely agrees with the more precise measurements performed at other scales.

\section{Particle distribution functions}

The measured scaling violations in particle distribution functions (PDFs) provide precise determinations of $\alpha_s$, taking advantage of the available deep-inelastic-scattering (DIS) data over a wide range of energies and, in particular, the accurate data sets obtained by the HERA experiments.

A combined analysis of non-singlet structure functions from DIS experiments \cite{Blumlein:2006be}, based on NNLO and N${}^3$LO QCD predictions, gave $\alpha_s(M_Z^2) = 0.1142\pm 0.0023$, including a theoretical error of $\pm 0.0008$ \cite{BDS:12}.
This determination neglects possible singlet contributions for $x> 0.35$, where the valence approximation is used.
More recent analyses take into account both singlet and non-singlet structure functions, together with Drell-Yan and di-muon data, needed for a correct description of the sea-quark densities, finding at NNLO the results
$\alpha_s(M_Z^2) = 0.1134\pm 0.0011$ \cite{Alekhin:2012ig}
and
$\alpha_s(M_Z^2) = 0.1158\pm 0.0035$ \cite{JimenezDelgado:2008hf}.

Global PDF analyses include a much broader set of data from fixed-target experiments, HERA and the Tevatron. In addition to scaling violations, the determination of the strong coupling obtained in this way exploits also the dependence on $\alpha_s$ of the hard-scattering matrix elements of the different processes analyzed. The inclusion of proton collider data allows for a better control of the gluon PDF. However, while full NNLO is employed for the structure functions, only NLO predictions are available for jet production at hadron colliders. The MSTW group finds $\alpha_s(M_Z^2) = 0.1171\pm 0.0014\pm 0.002$ \cite{Martin:2009bu}, where the last error indicates a conservatively estimated theoretical uncertainty. If the Tevatron data is removed from the fit, the central value decreases to $\alpha_s(M_Z^2) = 0.1104$; this DIS-only determination implies a gluon PDF incompatible with the Tevatron data, questioning the accuracy of previous DIS fits. In fact, fixing the high-$x$ gluon parameters to agree with the global PDF determination, the DIS-only fit gives a much larger value $\alpha_s(M_Z^2) = 0.1172$ \cite{Martin:2009bu}. Removing the old (but precise) BCDMS DIS data \cite{Benvenuti:1989fm}, the result from the global PDF fit increases to $\alpha_s(M_Z^2) = 0.1181$ while the DIS-only fit gives an even larger value $\alpha_s(M_Z^2) = 0.1193$ \cite{Martin:2009bu}. Pinning down the gluon PDF through Tevatron data seems to be crucial for a reliable $\alpha_s$ determination, at the expense of some deterioration in the fit quality of BCDMS data.

More recently, the NNPDF collaboration has performed a global PDF analysis, using neural networks as unbiased interpolating functions coupled to a Monte Carlo approach, with the result  $\alpha_s(M_Z^2) = 0.1173\pm 0.0011$
\cite{Ball:2011us}. Including only DIS data in the fit, they find also a smaller central value $\alpha_s(M_Z^2) = 0.1166$, in line with the MSTW results. They observe that at low $\alpha_s$ the DIS BCDMS data and jet data pull the gluon PDF in opposite directions, so the fit quality can be improved using DIS data only in a way which is forbidden when jet data are present.

As an educated weighted average of all these determinations, the PDG~\cite{BDS:12} quotes:
\begin{equation}\label{eq:pdf}
\alpha_s(M_Z^2)\; =\; 0.1151 \pm 0.0022\, .    
\end{equation}

The strong coupling has also been determined with inclusive jet measurements at HERA, over a wide range of values of the energy scale. 
The combined result, $\alpha_s(M_Z^2) = 0.1198\pm 0.0032$ \cite{Glasman:2007sm}, includes a large theoretical
uncertainty of $\pm 0.0026$. This determination has only a NLO accuracy and is not included in the average.

\section{Jet rates and event shapes in $e^+e^-$ annihilation}

Jet rates have a high sensitivity to $\alpha_s$, which increases with the jet multiplicity $n$ ($R_n\sim \alpha_s^{n-2}$). Moreover, there are many additional jet observables available, such as a variety of event shapes and energy correlations.
However, the physical description of jets always involves several scales ($E_{\mathrm{min}}$, $p_T$, $m_b$ \ldots), making necessary a careful resummation of enhanced logarithmic corrections and a good control of non-perturbative power corrections and hadronization effects.
Modern studies incorporate NNLO corrections \cite{GehrmannDeRidder:2009dp,Weinzierl:2010cw} and the most recent ones include a matched next-to-leading logarithmic (NLL) resummation \cite{Gehrmann:2008kh,Catani:1992ua}.
Using soft-collinear effective theory (SCET) techniques, this resummation has been achieved up to NNLL for jet broadening \cite{Becher:2012qc} and N${}^3$LL for
thrust \cite{Becher:2008cf} and heavy-jet mass \cite{Chien:2010kc}.

A NNLO analysis of the 3-jet rate, using ALEPH data collected at LEP between 91 and 209 GeV \cite{Heister:2003aj}, resulted in $\alpha_s(M_Z^2) = 0.1175\pm 0.0025$ \cite{Dissertori:2009qa}. The 5-jet rates at LEP have been also analyzed  at NLO with the result $\alpha_s(M_Z^2) = 0.1156\, {}^{+\, 0.0041}_{-\, 0.0034}$ \cite{Frederix:2010ne}. More recently, a re-analysis of the 3-jet JADE data, at centre-of-mass energies between 14 and 44 GeV, using NNLO predictions plus NLL resummation, obtains $\alpha_s(M_Z^2) = 0.1199\pm 0.0060$ \cite{Schieck:2012mp}, where the main uncertainty originates from applying different Monte Carlo models to estimate the transition from partons to hadrons. An older study of the 4-jet JADE data, at NLO plus NLL, gave
$\alpha_s(M_Z^2) = 0.1159\pm 0.0028$ \cite{Schieck:2006tc}.

Two analyses of six event-shape variables with LEP data, performed at NNLO plus NLL, have given the values $\alpha_s(M_Z^2) = 0.1224\pm 0.0039$ \cite{Dissertori:2009ik} and $\alpha_s(M_Z^2) = 0.1189\pm 0.0043$ \cite{OPAL:2011aa}. In both cases, the dominant uncertainty is theoretical. A similar event-shape analysis of JADE data gives $\alpha_s(M_Z^2) = 0.1172\pm 0.0051$, where hadronization ($\pm 0.0035$) and perturbative QCD ($\pm 0.0030$) are the largest errors \cite{Bethke:2008hf}.
Including a N${}^3$LL resummation into the NNLO predictions, a fit to the thrust data from ALEPH and OPAL
leads to $\alpha_s(M_Z^2) = 0.1172 \pm 0.0022$ \cite{Becher:2008cf}. Using only ALEPH data and working also at
NNLO plus N${}^3$LL, Ref.~\cite{Chien:2010kc} finds a similar result from thrust and a larger value from the heavy-jet mass distribution, with a final combined result $\alpha_s(M_Z^2) = 0.1193 \pm 0.0027$. Ref.~\cite{Becher:2008cf} includes a $\pm 0.0012$ hadronization uncertainty, while hadronization is neglected in Ref.~\cite{Chien:2010kc}.

In the previous analyses hadronization effects are estimated with Monte Carlo generators, which were tuned to LEP data. Higher-order perturbative contributions, non-perturbative power corrections and hadron-mass effects get then unavoidably mixed, which can have a significant impact on the $\alpha_s$ extraction.
Using SCET techniques, Refs.~\cite{Abbate:2012jh} incorporate explicitly the leading non-perturbative power corrections, which are also fitted to the data. They analyse the world data on thrust distributions at NNLO plus N${}^3$LL accuracy, including a sophisticated infrared renormalon subtraction, obtaining quite low values of the strong coupling. The tail of the thrust distribution gives $\alpha_s(M_Z^2) = 0.1135\pm 0.0011$, while cumulant moments using the full thrust distribution lead to
$\alpha_s(M_Z^2) = 0.1140\pm 0.0015$ \cite{Abbate:2012jh}.
In both cases, the inclusion in the fit of an inverse power correction results in a large decrease of the central value ($\Delta\alpha_s\sim -0.009$), while the total uncertainty has been reduced by a factor close to 3 after the renormalon subtraction. The size of subleading power corrections, not included in the fit, remains to be investigated.

A small $\alpha_s$ value with a larger uncertainty,
$\alpha_s(M_Z^2) = 0.1131\, {}^{+\, 0.0028}_{-\, 0.0022}$,
has been also obtained in another thrust analysis at NNLO plus NNLL \cite{Gehrmann:2012sc}, where hadronization effects are taken into account analytically through an effective coupling frozen in the infrared.
A previous thrust analysis within the same hadronization approach at NNLO plus NLL found
$\alpha_s(M_Z^2) = 0.1164\, {}^{+\, 0.0034}_{-\, 0.0032}$ \cite{Davison:2008vx}.
This approach has also been applied at NNLO to moments of other event-shape variables with the result
$\alpha_s(M_Z^2) = 0.1153\pm 0.0029$ \cite{Gehrmann:2009eh}.

\begin{figure}[t]\centering
\includegraphics[width=.48\textwidth]{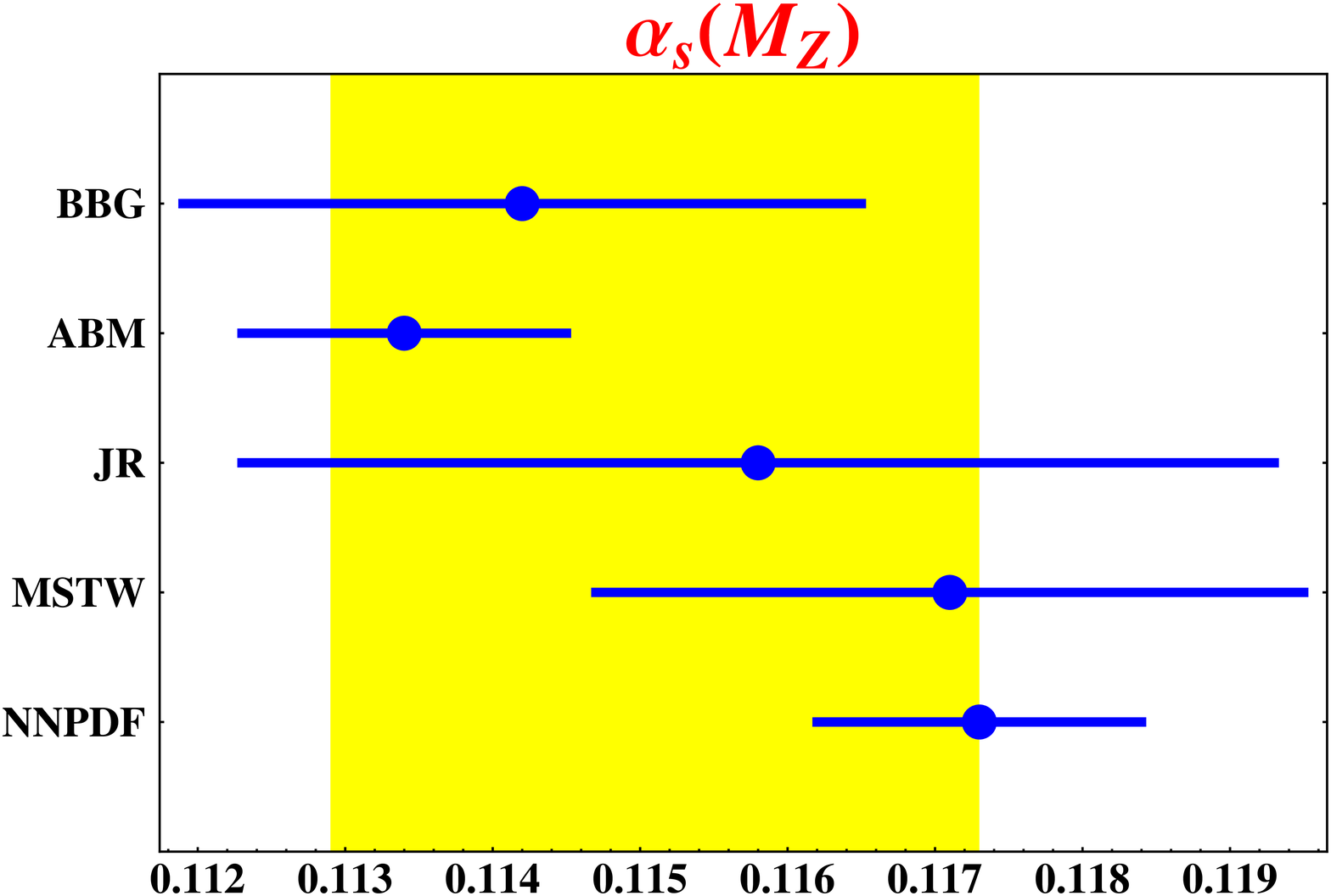}
\hskip .4cm
\includegraphics[width=.48\textwidth]{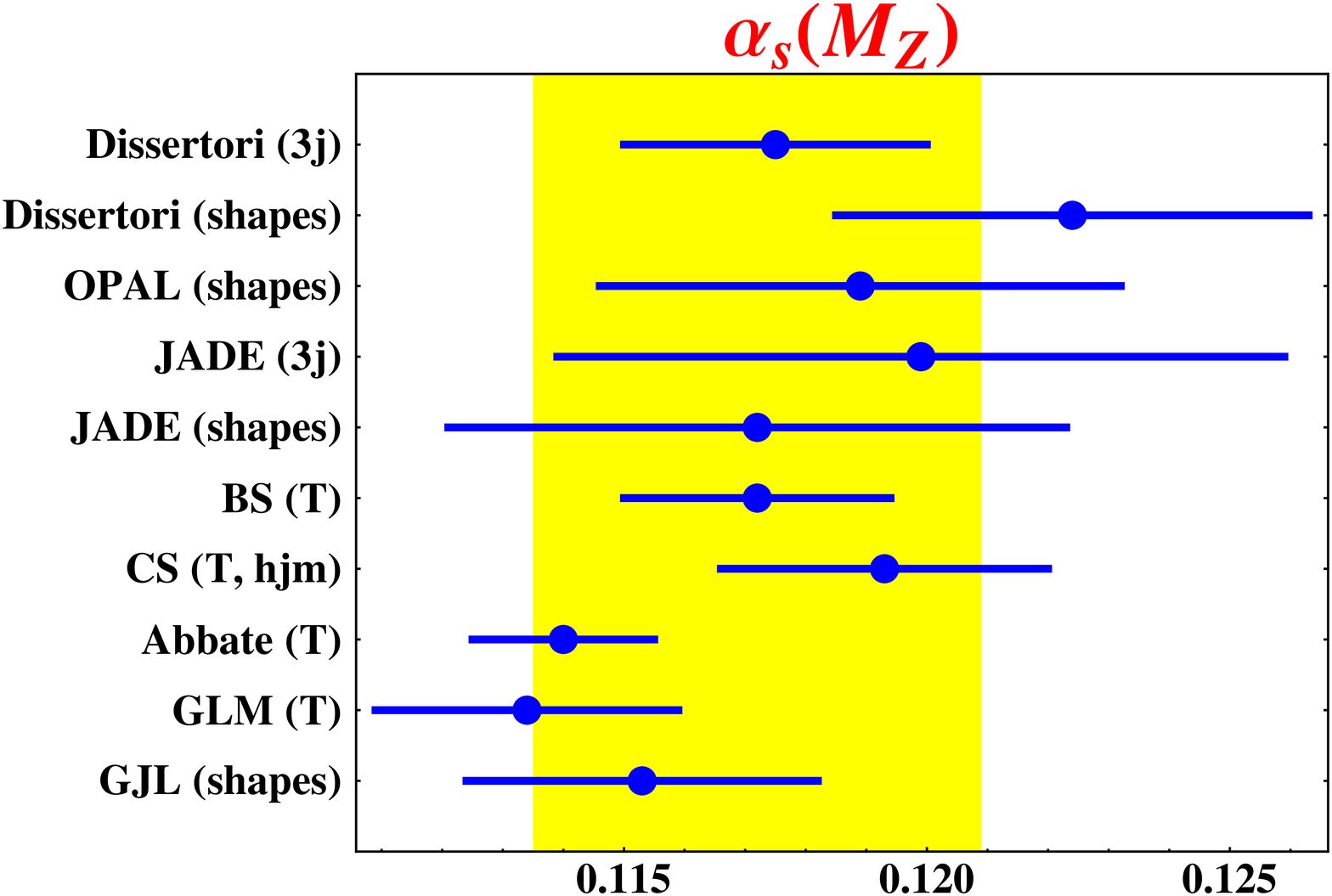} 
\caption{
$\alpha_s(M_Z^2)$ from PDFs (left) and $e^+e^-$ jets (right).
The yellow bands indicate the PDG \cite{BDS:12} averages.}
\label{fig:pre-averages}
\end{figure}

As an educated average of $e^+e^-$ results based on NNLO predictions, the PDG quotes \cite{BDS:12}
\begin{equation}\label{eq:jets}
\alpha_s(M_Z^2)\; =\; 0.1172 \pm 0.0037\, .
\end{equation}

\section{Jets at hadron colliders}

The D0 collaboration extracted the strong coupling from the $p_T$ dependence of the inclusive jet cross section in $p\bar p$ collisions at $\sqrt{s}=1.96$~TeV, which is known to NLO, with the result
$\alpha_s(M_Z^2) = 0.1161\, {}^{+\, 0.0041}_{-\, 0.0048}$  \cite{Abazov:2009nc}. The uncertainty is dominated by experimental errors (jet energy calibration, $p_T$ resolution and integrated luminosity).
More recently, from a NLO study of jet angular correlations over a wide range of momentum transfers from 50 to 400 GeV, D0 obtains $\alpha_s(M_Z^2) = 0.1191\, {}^{+\, 0.0048}_{-\, 0.0071}$, finding good agreement
with the predicted QCD running up to 400 GeV \cite{Abazov:2012lua}.

A first NLO determination based on LHC data was performed with the inclusive jet cross section measured by ATLAS at $\sqrt{s}= 7$~TeV, testing the running of $\alpha_s$ up to jet $p_T$ of 600~GeV, with the result
$\alpha_s(M_Z^2) = 0.1151\, {}^{+\, 0.0093}_{-\, 0.0087}$ \cite{Malaescu:2012ts}.
CMS has made public two preliminary results:
the measured ratio of the inclusive 3-jet and 2-jet cross sections at $\sqrt{s}= 7$~TeV,
as a function of the average transverse momentum of the two leading jets up to 1.4~TeV, gives
$\alpha_s(M_Z^2) = 0.1143\, {}^{+\, 0.0083}_{-\, 0.0067}$ \cite{CMS:2013}, while the
comparison of the $t\bar t$ production cross section with approximate NNLO calculations results in
$\alpha_s(M_Z^2) = 0.1178\, {}^{+\, 0.0046}_{-\, 0.0040}$ \cite{CMS:2013b}.

\section{Summary: the world average value of $\alpha_s$}

Making a world average over the different determinations of $\alpha_s$ is a highly non-trivial task because systematic uncertainties dominate the most precise measurements. Thus, one relies in the more or less conservative attitude adopted to estimate the errors of a given determination. Different levels of theoretical accuracy exist for the different observables analyzed and, moreover, many theoretical and experimental inputs are highly correlated.
Following the procedure suggested in Refs.~\cite{Bethke:2009jm,Bethke:2011tr,BDS:12}, we just combine the pre-averages given in the previous sections for each class of measurements, which only take into account determinations of at least NNLO accuracy. The weighted average of the different input values gives the central value, while the uncertainty is adjusted so that the $\chi^2/\mathrm{dof}$ equals unity. One finds in this way
\begin{equation}\label{eq:average}
\alpha_s(M_Z^2)\; =\; 0.1186 \pm 0.0007\, ,
\end{equation}
which is very close to the value quoted in the 2012 PDG review \cite{BDS:12}. As nicely discussed there, the central value remains stable when removing any one of the five inputs included in this average. The overall uncertainty is however largely determined by the precise lattice result.

\begin{figure}[t]\centering
\includegraphics[width=.8\textwidth]{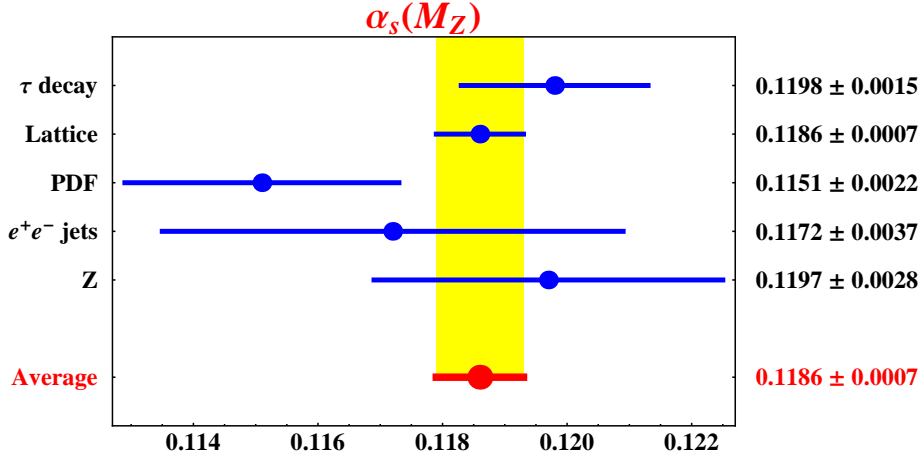}
\caption{Summary of $\alpha_s(M_Z^2)$ determinations.}
\label{fig:summary}
\end{figure}

\section*{Acknowledgements}
I want to thank the organizers for this enjoyable conference and V. Mateu and P. Ruiz for useful discussions.
This work has been supported by the Spanish Government
and EU ERDF funds 
[grants FPA2007-60323, FPA2011-23778 and CSD2007-00042 (Consolider Project CPAN)].


\begin{thebibliography}{99}

\bibitem{Bethke:2009jm}
  S.~Bethke,
  Eur.\ Phys.\ J.\ C {\bf 64} (2009) 689;
%
arXiv:1210.0325 [hep-ex].

\bibitem{Bethke:2011tr}
  S.~Bethke {\it et al.},  
  arXiv:1110.0016 [hep-ph].

\bibitem{BDS:12}
S. Bethke, G. Dissertori and G.~P. Salam, 
in \cite{Beringer:1900zz}, p.~120.

\bibitem{Beringer:1900zz}
  J.~Beringer {\it et al.}, 
  Phys.\ Rev.\ D {\bf 86} (2012) 010001.

\bibitem{vanRitbergen:1997va}
  T.~van Ritbergen, J.~A.~M.~Vermaseren and S.~A.~Larin,
  Phys.\ Lett.\ B {\bf 400} (1997) 379.

\bibitem{Czakon:2004bu}
  M.~Czakon,
  Nucl.\ Phys.\ B {\bf 710} (2005) 485.


\bibitem{Schroder:2005hy}
  Y.~Schroder and M.~Steinhauser,
  JHEP {\bf 0601} (2006) 051.

\bibitem{Chetyrkin:2005ia}
  K.~G.~Chetyrkin, J.~H.~Kuhn and C.~Sturm,
  Nucl.\ Phys.\ B {\bf 744} (2006) 121.

\bibitem{Baikov:2008jh}
  P.~A.~Baikov, K.~G.~Chetyrkin and J.~H.~Kuhn,
  Phys.\ Rev.\ Lett.\  {\bf 101} (2008) 012002.


\bibitem{Baikov:2012er}
  P.~A.~Baikov {\it et al.},  
  Phys.\ Rev.\ Lett.\  {\bf 108} (2012) 222003;
%
  Phys.\ Lett.\ B {\bf 714} (2012) 62.

\bibitem{Gorishnii:1990vf}
  S.~G.~Gorishnii, A.~L.~Kataev and S.~A.~Larin,
  Phys.\ Lett.\ B {\bf 259} (1991) 144.

\bibitem{Surguladze:1990tg}
  L.~R.~Surguladze and M.~A.~Samuel,
  Phys.\ Rev.\ Lett.\  {\bf 66} (1991) 560
   [Erratum-ibid.\  {\bf 66} (1991) 2416].


\bibitem{Narison:1988ni}
  S.~Narison and A.~Pich,
  Phys.\ Lett.\ B {\bf 211} (1988) 183.

\bibitem{Braaten:1988hc}
  E.~Braaten,
  Phys.\ Rev.\ Lett.\  {\bf 60} (1988) 1606;
%
  Phys.\ Rev.\ D {\bf 39} (1989) 1458.

\bibitem{Braaten:1991qm}
  E.~Braaten, S.~Narison and A.~Pich,
  Nucl.\ Phys.\ B {\bf 373} (1992) 581.

\bibitem{Marciano:1988vm}
  W.~J.~Marciano and A.~Sirlin,
  Phys.\ Rev.\ Lett.\  {\bf 61} (1988) 1815.

\bibitem{Braaten:1990ef}
  E.~Braaten and C.~-S.~Li,
  Phys.\ Rev.\ D {\bf 42} (1990) 3888.

\bibitem{Erler:2002mv}
  J.~Erler,
  Rev.\ Mex.\ Fis.\  {\bf 50} (2004) 200.

\bibitem{Le Diberder:1992fr}
  F.~Le Diberder and A.~Pich,
  Phys.\ Lett.\ B {\bf 289} (1992) 165.

\bibitem{Schael:2005am} ALEPH Collaboration,
  Phys.\ Rept.\  {\bf 421} (2005) 191;
%
  Eur.\ Phys.\ J.\ C {\bf 4} (1998) 409;
%
  Phys.\ Lett.\ B {\bf 307} (1993) 209.

\bibitem{Ackerstaff:1998yj} OPAL Collaboration,
  Eur.\ Phys.\ J.\ C {\bf 7} (1999) 571.

\bibitem{Coan:1995nk} CLEO Collaboration,
  Phys.\ Lett.\ B {\bf 356} (1995) 580.

\bibitem{Davier:2008sk}
  M.~Davier {\it et al.}, 
  Eur.\ Phys.\ J.\ C {\bf 56} (2008) 305;
%
  Rev.\ Mod.\ Phys.\  {\bf 78} (2006) 1043.


\bibitem{Pich:1998yn}
  A.~Pich and J.~Prades,
  JHEP {\bf 9806} (1998) 013,
%
  {\bf 9910} (1999) 004.

\bibitem{Baikov:2004tk}
  P.~A.~Baikov, K.~G.~Chetyrkin and J.~H.~Kuhn,
  Phys.\ Rev.\ Lett.\  {\bf 95} (2005) 012003.

\bibitem{Le Diberder:1992te}
  F.~Le Diberder and A.~Pich,
  Phys.\ Lett.\ B {\bf 286} (1992) 147.


\bibitem{Pich:2011bb}
  A.~Pich,
  arXiv:1107.1123 [hep-ph];
  Nucl.\ Phys.\ Proc.\ Suppl.\  {\bf 218} (2011) 89;
  Acta Phys.\ Polon.\ Supp.\  {\bf 3} (2010) 165.

\bibitem{Beneke:2008ad}
  M.~Beneke and M.~Jamin,
  JHEP {\bf 0809} (2008) 044.
%
  M.~Beneke, D.~Boito and M.~Jamin,
  JHEP {\bf 1301} (2013) 125.

\bibitem{Caprini:2011ya}
  I.~Caprini and J.~Fischer,
  Phys.\ Rev.\ D {\bf 84} (2011) 054019;
%
  Eur.\ Phys.\ J.\ C {\bf 64} (2009) 35.

\bibitem{Abbas:2012fi}
  G.~Abbas, B.~Ananthanarayan, I.~Caprini and J.~Fischer,
  Phys.\ Rev.\ D {\bf 87} (2013) 014008.

\bibitem{Abbas:2012py}
  G.~Abbas, B.~Ananthanarayan and I.~Caprini,
  Phys.\ Rev.\ D {\bf 85} (2012) 094018.

\bibitem{Groote:2012jq}
  S.~Groote, J.~G.~Korner and A.~A.~Pivovarov,
  arXiv:1212.5346 [hep-ph].

\bibitem{Maltman:2008nf}
  K.~Maltman and T.~Yavin,
  Phys.\ Rev.\ D {\bf 78} (2008) 094020.

\bibitem{Menke:2009vg}
  S.~Menke,
  arXiv:0904.1796 [hep-ph].

\bibitem{Narison:2009vy}
  S.~Narison,
  Phys.\ Lett.\ B {\bf 673} (2009) 30.

\bibitem{Cvetic:2010ut}
  G.~Cvetic, M.~Loewe, C.~Martinez and C.~Valenzuela,
  Phys.\ Rev.\ D {\bf 82} (2010) 093007.

\bibitem{Pivovarov:1991rh}
  A.~A.~Pivovarov,
  Z.\ Phys.\ C {\bf 53} (1992) 461.


\bibitem{HFAG} Heavy Flavor Averaging Group, 
arXiv:1207.1158 [hep-ex];
http://www.slac.stanford.edu/xorg/hfag/.



\bibitem{DescotesGenon:2010cr}
  S.~Descotes-Genon and B.~Malaescu,
  arXiv:1002.2968 [hep-ph].

\bibitem{Cata:2008ru}
  O.~Cata, M.~Golterman and S.~Peris,
  Phys.\ Rev.\ D {\bf 79} (2009) 053002,
  {\bf 77} (2008) 093006.

\bibitem{Boito:2012cr}
  D.~Boito {\it et al.},
  Phys.\ Rev.\ D {\bf 85} (2012) 093015,
%
  {\bf 84} (2011) 113006.

\bibitem{McNeile:2010ji}
  C.~McNeile {\it et al.},     
  Phys.\ Rev.\ D {\bf 82} (2010) 034512.
%
  C.~T.~H.~Davies {\it et al.}, 
  Phys.\ Rev.\ D {\bf 78} (2008) 114507.

\bibitem{Mason:2005zx}
  Q.~Mason {\it et al.},  
  Phys.\ Rev.\ Lett.\  {\bf 95} (2005) 052002.

\bibitem{Maltman:2008bx}
  K.~Maltman, D.~Leinweber, P.~Moran and A.~Sternbeck,
  Phys.\ Rev.\ D {\bf 78} (2008) 114504.

\bibitem{Aoki:2009tf}
  S.~Aoki {\it et al.},  
  JHEP {\bf 0910} (2009) 053.

\bibitem{DellaMorte:2004bc}
  M.~Della Morte {\it et al.},   
  Nucl.\ Phys.\ B {\bf 713} (2005) 378.
%
  R.~Sommer, F.~Tekin and U.~Wolff,
  PoS LATTICE {\bf 2010} (2010) 241.

\bibitem{Shintani:2010ph}
  E.~Shintani {\it et al.},   
  Phys.\ Rev.\ D {\bf 82} (2010) 074505.

\bibitem{Blossier:2012ef}
  B.~Blossier {\it et al.}, 
  Phys.\ Rev.\ Lett.\  {\bf 108} (2012) 262002.

\bibitem{RBC:12}
R. J. Hudspith, P. Boyle and L. Del Debbio, talk at 
the 30th International Symposium on Lattice Field Theory
(Cairns, Australia, 2012).

\bibitem{Bazavov:2011nk}
  A.~Bazavov {\it et al.},  
  Phys.\ Rev.\ D {\bf 85} (2012) 054503.

\bibitem{Brambilla:2010pp}
  N.~Brambilla, X.~Garcia i Tormo, J.~Soto and A.~Vairo,
  Phys.\ Rev.\ Lett.\  {\bf 105} (2010) 212001
   [Erratum-ibid.\  {\bf 108} (2012) 269903];
%
  Phys.\ Rev.\ D {\bf 80} (2009) 034016;
%
  Phys.\ Lett.\ B {\bf 647} (2007) 185.

\bibitem{Pineda:2011db}
  A.~Pineda and M.~Stahlhofen,
  Phys.\ Rev.\ D {\bf 84} (2011) 034016.

\bibitem{Smirnov:2009fh}
  A.~V.~Smirnov, V.~A.~Smirnov and M.~Steinhauser,
  Phys.\ Rev.\ Lett.\  {\bf 104} (2010) 112002;
%
  Phys.\ Lett.\ B {\bf 668} (2008) 293.

\bibitem{Anzai:2009tm}
  C.~Anzai, Y.~Kiyo and Y.~Sumino,
  Phys.\ Rev.\ Lett.\  {\bf 104} (2010) 112003.

\bibitem{Bazavov:2012ka}
  A.~Bazavov {\it et al.},  
  Phys.\ Rev.\ D {\bf 86} (2012) 114031.

\bibitem{Besson:2005jv} CLEO Collaboration,
  Phys.\ Rev.\ D {\bf 74} (2006) 012003.

\bibitem{GarciaiTormo:2005ch}
  X.~Garcia i Tormo and J.~Soto,
  Phys.\ Rev.\ D {\bf 72} (2005) 054014.



\bibitem{Brambilla:2007cz}
  N.~Brambilla, X.~Garcia i Tormo, J.~Soto and A.~Vairo,
  Phys.\ Rev.\ D {\bf 75} (2007) 074014.



\bibitem{Blumlein:2006be}
  J.~Blumlein, H.~Bottcher and A.~Guffanti,
  Nucl.\ Phys.\ B {\bf 774} (2007) 182.

\bibitem{Alekhin:2012ig}
  S.~Alekhin, J.~Blumlein and S.~Moch,
  Phys.\ Rev.\ D {\bf 86} (2012) 054009.
%
  S.~Alekhin, J.~Blumlein, S.~Klein and S.~Moch,
  Phys.\ Rev.\ D {\bf 81} (2010) 014032.

\bibitem{JimenezDelgado:2008hf}
  P.~Jimenez-Delgado and E.~Reya,
  Phys.\ Rev.\ D {\bf 79} (2009) 074023.

\bibitem{Martin:2009bu}
  A.~D.~Martin, W.~J.~Stirling, R.~S.~Thorne and G.~Watt,
  Eur.\ Phys.\ J.\ C {\bf 64} (2009) 653.

\bibitem{Benvenuti:1989fm} BCDMS Collaboration,
  Phys.\ Lett.\ B {\bf 237} (1990) 592,
%
{\bf 223} (1989) 485.

\bibitem{Ball:2011us}
  R.~D.~Ball {\it et al.},  
  Phys.\ Lett.\ B {\bf 707} (2012) 66,
%
  {\bf 701} (2011) 346.

\bibitem{Glasman:2007sm}
  C.~Glasman [H1 and ZEUS Collaborations],
  J.\ Phys.\ Conf.\ Ser.\  {\bf 110} (2008) 022013.


\bibitem{GehrmannDeRidder:2009dp}
  A.~Gehrmann-De Ridder, T.~Gehrmann, E.~W.~N.~Glover and G.~Heinrich,
  JHEP {\bf 0905} (2009) 106,
%
  {\bf 0712} (2007) 094,
%
  {\bf 0711} (2007) 058;
%
  Phys.\ Rev.\ Lett.\  {\bf 100} (2008) 172001,
%
  {\bf 99} (2007) 132002.

\bibitem{Weinzierl:2010cw}
  S.~Weinzierl,
  Eur.\ Phys.\ J.\ C {\bf 71} (2011) 1565
   [Erratum-ibid.\ C {\bf 71} (2011) 1717];
%
  Phys.\ Rev.\ D {\bf 80} (2009) 094018;
%
  JHEP {\bf 0906} (2009) 041;
%
  Phys.\ Rev.\ Lett.\  {\bf 101} (2008) 162001.
%

\bibitem{Gehrmann:2008kh}
  T.~Gehrmann, G.~Luisoni and H.~Stenzel,
  Phys.\ Lett.\ B {\bf 664} (2008) 265.

\bibitem{Catani:1992ua}
  S.~Catani, L.~Trentadue, G.~Turnock and B.~R.~Webber,
  Nucl.\ Phys.\ B {\bf 407} (1993) 3.

\bibitem{Becher:2012qc}
  T.~Becher and G.~Bell,
  JHEP {\bf 1211} (2012) 126.

\bibitem{Becher:2008cf}
  T.~Becher and M.~D.~Schwartz,
  JHEP {\bf 0807} (2008) 034.

\bibitem{Chien:2010kc}
  Y.~-T.~Chien and M.~D.~Schwartz,
  JHEP {\bf 1008} (2010) 058.


\bibitem{Heister:2003aj} ALEPH Collaboration,
  Eur.\ Phys.\ J.\ C {\bf 35} (2004) 457.

\bibitem{Dissertori:2009qa}
  G.~Dissertori {\it et al.},  
  Phys.\ Rev.\ Lett.\  {\bf 104} (2010) 072002.

\bibitem{Schieck:2012mp} JADE Collaboration,
  arXiv:1205.3714 [hep-ex].

\bibitem{Schieck:2006tc} JADE Collaboration,
  Eur.\ Phys.\ J.\ C {\bf 48} (2006) 3
   [Erratum-ibid.\ C {\bf 50} (2007) 769].

\bibitem{Frederix:2010ne}
  R.~Frederix, S.~Frixione, K.~Melnikov and G.~Zanderighi,
  JHEP {\bf 1011} (2010) 050.

\bibitem{Dissertori:2009ik}
  G.~Dissertori {\it et al.}, 
  JHEP {\bf 0908} (2009) 036,
%
  {\bf 0802} (2008) 040.

\bibitem{OPAL:2011aa} OPAL Collaboration,
  Eur.\ Phys.\ J.\ C {\bf 71} (2011) 1733.

\bibitem{Bethke:2008hf} JADE Collaboration,
  Eur.\ Phys.\ J.\ C {\bf 64} (2009) 351.

\bibitem{Abbate:2012jh}
  R.~Abbate {\it et al.},  
  Phys.\ Rev.\ D {\bf 86} (2012) 094002,
%
  {\bf 83} (2011) 074021.

\bibitem{Gehrmann:2012sc}
  T.~Gehrmann, G.~Luisoni and P.~F.~Monni,
  Eur.\ Phys.\ J.\ C {\bf 73} (2013) 2265;
%
  JHEP {\bf 1108} (2011) 010.

\bibitem{Davison:2008vx}
  R.~A.~Davison and B.~R.~Webber,
  Eur.\ Phys.\ J.\ C {\bf 59} (2009) 13;
and update in \cite{Bethke:2011tr}.

\bibitem{Gehrmann:2009eh}
  T.~Gehrmann, M.~Jaquier and G.~Luisoni,
  Eur.\ Phys.\ J.\ C {\bf 67} (2010) 57.


\bibitem{Abazov:2009nc} D0 Collaboration,
  Phys.\ Rev.\ D {\bf 80} (2009) 111107.

\bibitem{Abazov:2012lua} D0 Collaboration,
  Phys.\ Lett.\ B {\bf 718} (2012) 56.

\bibitem{Malaescu:2012ts}
  B.~Malaescu and P.~Starovoitov,
  Eur.\ Phys.\ J.\ C {\bf 72} (2012) 2041.

\bibitem{CMS:2013} CMS Collaboration, CMS-PAS-QCD-11-003 (2012).

\bibitem{CMS:2013b} CMS Collaboration, CMS-PAS-TOP-12-022 (2012).

\end{thebibliography}
\end{document}